\documentclass[preprint,prl,amsmath,amssymb]{revtex4-1}

\usepackage{graphicx}
\usepackage{mathptmx}

\usepackage{color}
\usepackage{ulem}

\begin{document}

\title{Structural and electronic properties of epitaxial multilayer \textit{h}-BN on Ni(111) for spintronics applications}

\author{A. A. Tonkikh,$^{1,2}$ E. N. Voloshina,$^3$ P. Werner,$^1$ H. Blumtritt,$^1$ B. Senkovskiy,$^{4,5}$ G. G\"untherodt,$^{1,6}$ S. S. P. Parkin,$^1$ and Yu. S. Dedkov$^{7,}$\footnote{Corresponding author. E-mail: Yuriy.Dedkov@specs.com; Yuriy.Dedkov@icloud.com}}

\affiliation{\mbox{$^1$Max Planck Institute of Microstructure Physics, Weinberg 2, 06120 Halle (Saale), Germany}}
\affiliation{\mbox{$^2$Institute for Physics of Microstructures RAS, 603950, GSP-105, Nizhny Novgorod, Russia}}
\affiliation{\mbox{$^3$Humboldt-Universit\"at zu Berlin, Institut f\"ur Chemie, 10099 Berlin, Germany}}
\affiliation{\mbox{$^4$Institute of Solid State Physics, Dresden University of Technology, 01062 Dresden, Germany}}
\affiliation{\mbox{$^5$St. Petersburg State University, 198504 St. Petersburg, Russia}}
\affiliation{\mbox{$^6$2nd Institute of Physics and JARA-FIT, RWTH Aachen University, 52074 Aachen, Germany}}
\affiliation{\mbox{$^7$SPECS Surface Nano Analysis GmbH, Voltastra\ss e 5, 13355 Berlin, Germany}}

\date{\today}

\begin{abstract} 
Hexagonal boron nitride ($h$-BN) is a promising material for implementation in spintronics due to a large band gap, low spin-orbit coupling, and a small lattice mismatch to graphene and to close-packed surfaces of fcc-Ni(111) and hcp-Co(0001). Epitaxial deposition of $h$-BN on ferromagnetic metals is aimed at small interface scattering of charge and spin carriers. We report on the controlled growth of $h$-BN/Ni(111) by means of molecular beam epitaxy (MBE). Structural and electronic properties of this system are investigated using cross-section transmission electron microscopy (TEM) and electron spectroscopies which confirm good agreement with the properties of bulk $h$-BN. The latter are also corroborated by density functional theory (DFT) calculations, revealing that the first $h$-BN layer at the interface to Ni is metallic. Our investigations demonstrate that MBE is a promising, versatile alternative to both the exfoliation approach and chemical vapour deposition of $h$-BN.
\end{abstract}

\maketitle

\section*{Introduction}

The fast development of the modern information technology raises the quest for advanced materials and new methods of nano-devices fabrication.  A prominent topical example is graphene, a two-dimensional (2D) crystalline atomic layer system with charge carrier mobilities exceeding by far those of Si~\cite{Novoselov:2004a}. This ideal two-dimensional electron gas (2DEG) system exhibits unique properties for charge~\cite{Novoselov:2005,Zhang:2005} and spin~\cite{Tombros:2007} transport. The spin property prompts applications in memory, sensor, and active spin-current based devices. For the latter the detrimental spin transport parameters like spin relaxation time and relaxation length have called for materials with negligible intrinsic spin-orbit and hyperfine interactions. Here graphene appears as a prime spintronics candidate being gate tunable at room temperature~\cite{Tombros:2007,Yang:2011,Han:2011id}. Thereby, graphene has also spurred the conjectured change in paradigm from charge-based electronics, i.\,e. transfer and storage of electron charges, to spintronics~\cite{Bader:2010gr} based on the electron's spin.

The progress in graphene-based devices~\cite{Geim:2007a,CastroNeto:2009,Novoselov:2011} triggered the search for a suitable insulating substrate. The crystalline structure of wide band gap hexagonal boron nitride ($h$-BN) is isomorphic to the structure of graphite making $h$-BN an ideal match to graphene. Weak van der Waals bonds between honeycomb $h$-BN sheets and the absence of surface states allow easy manipulations with exfoliated single or multilayers of $h$-BN. Owing to these properties new physical phenomena~\cite{Amet:2015ka,Faugeras:2015it,Natterer:2015kv,Kumar:2015ct} and new devices~\cite{Gao:2015jl,Gaskell:2015io,Parzefall:2015eo,Petrone:2015kk} became possible with exfoliated $h$-BN. In particular, superior charge and spin transport properties in graphene on $h$-BN~\cite{Banszerus:2015bx,Drogeler:2014cw} over graphene on amorphous SiO$_2$~\cite{Kamalakar:2015is} are reported. In this case, the smoothening of graphene flakes by $h$-BN flakes has enabled significant improvements of spin lifetimes, spin relaxation lengths, and electron mobilities in Co/MgO/graphene/$h$-BN spin-valve devices~\cite{Drogeler:2014cw}.

Moreover, for spin injection into graphene it has been realized that $h$-BN with a large band gap of $\approx6$\,eV~\cite{Tarrio:1989ty,Kubota:2007jl} has the potential of replacing  defective oxide tunnel barriers~\cite{Lee:2011cf,Britnell:2013ca}. In view of early predictions by Kirczenow~\cite{Kirczenow:2001ef} for metal/$h$-BN/semiconductor interfaces the analogous barrier replacement could be assumed for magnetic tunnel junctions (MTJs). Currently, MTJs are made of ferromagnetic electrodes separated by Al$_2$O$_3$ or a hygroscopic MgO barrier~\cite{Parkin:2004do}. An attempt of making MTJs with  CVD grown $h$-BN tunnel barriers has resulted in a rather low tunneling magnetoresistance (TMR) of less than $1\%$~\cite{Dankert:2014ip}. Such TMR is far below predicted values~\cite{Yazyev:2009} due to contaminations during $h$-BN transfer.

All these examples of excellent physical  properties and  specific device demonstrations by applying exfoliated $h$-BN are preceding the next stage in the development of van der Waals materials: large scale and integrated device fabrication, requiring  wafer scale fabrication methods of $h$-BN. Recently, several approaches of epitaxial $h$-BN growth on close-packed surfaces of polycrystalline transition metals have been proposed including borasine-free chemical vapour deposition~\cite{Ismach:2012hn,Lu:2015ev}, sputtering~\cite{Sutter:2013eu}, and molecular beam epitaxy (MBE)~\cite{Nakhaie:2015vl}. In these publications mainly structural properties of epitaxial $h$-BN films have been addressed, while their electronic structures have not been reported yet. Our work addresses this topic by presenting a comprehensive characterization of large scale epitaxial h-BN films complemented by density-functional theory (DFT) modeling.

We report on the epitaxial growth of multilayered $h$-BN (ml-$h$-BN) on the close-packed Ni(111) surface of single crystal Ni by means of MBE. The obtained films of ml-$h$-BN demonstrate very high structural perfection with A-B stacking of BN layers as observed by means of cross-sectional transmission electron microscopy (TEM). Photoelectron spectroscopy experiments demonstrate clearly the energy dispersion of the valence band states below the Fermi level ($E_F$) of the system. These data in combination with near-edge x-ray absorption spectroscopy (NEXAFS) results and DFT calculations allow for the systematic mapping of the electronic structure of the system. Our findings demonstrate that MBE growth of ml-$h$-BN is a perspective method that can be easily adapted to electronic and spintronic applications of $h$-BN such as, e.\,g., advanced devices for spin injection into graphene or for MTJs.

\section*{Results}

\begin{figure}
\includegraphics[scale=0.2]{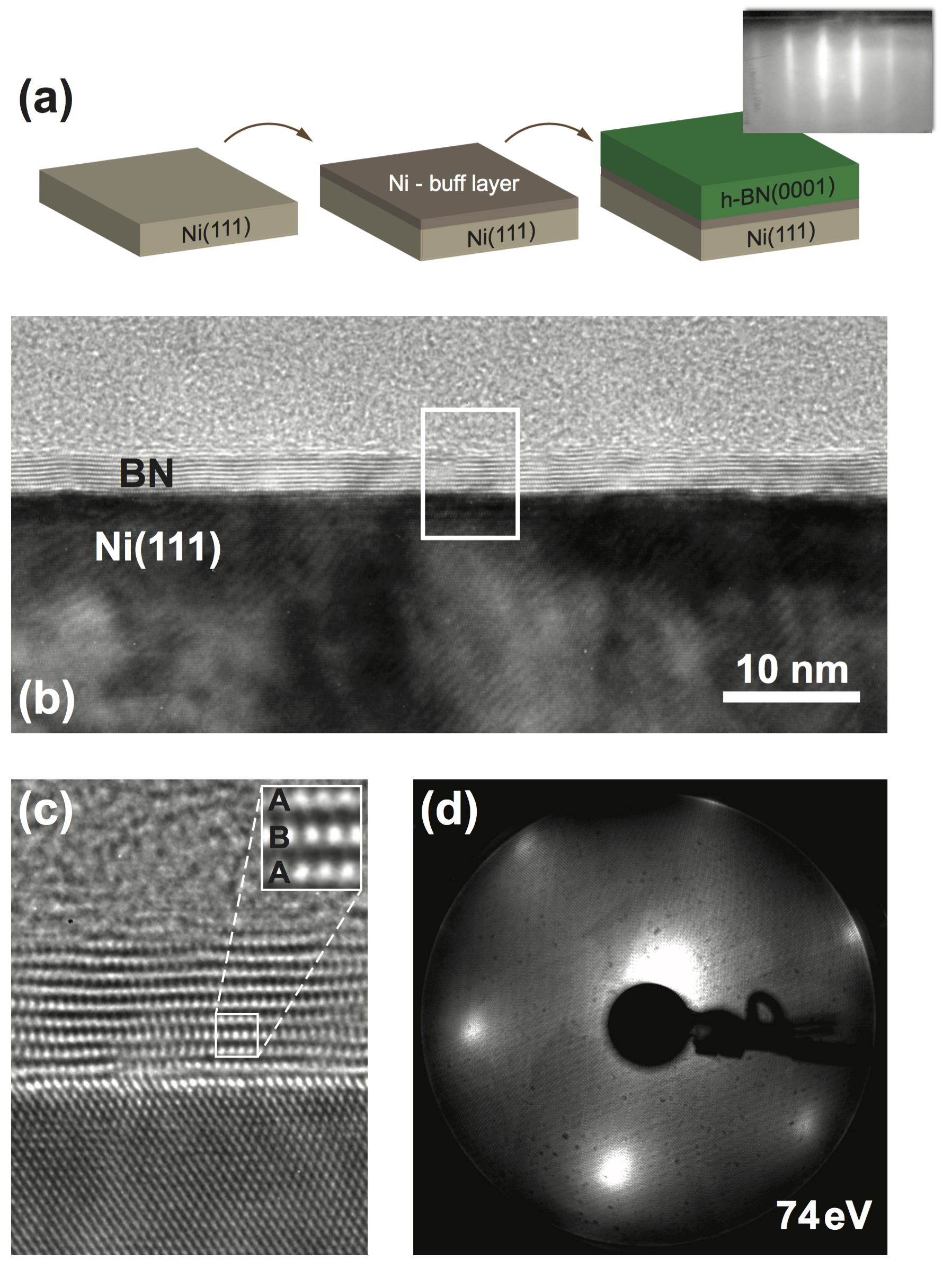}\\
\caption{\label{TEM-LEED} \textbf{Structural characterization of h-BN/Ni(111).} (a) Step-by-step preparation of the $h$-BN/Ni(111) structures by MBE. RHEED image of this system is shown as an inset. (b,c) High-resolution TEM images of the $10$\,ML-thick BN on Ni(111). Inset in (c) shows a zoomed image of the ml-$h$-BN demonstrating the A-B stacking of BN layers. (d) LEED image of the same sample obtained after atmospheric transfer of ml-$h$-BN/Ni(111) into a UHV chamber and annealing at $400^\circ$\,C. The energy of the primary electron beam is $74$\,eV.} 
\end{figure}

\textbf{$h$-BN/Ni(111) growth and characterization.} The epitaxial ml-$h$-BN films were grown using the MBE technique [Fig.~\ref{TEM-LEED}(a)], by which boron atoms were deposited in the presence of nitrogen atoms, generated in an RF-plasma source, on Ni(111) substrates kept at $700^\circ$\,C (for details, see section Methods). The results of the structural characterization of the ml-$h$-BN/Ni(111) stack are compiled in Fig.~\ref{TEM-LEED} for a 10\,ML-thick $h$-BN film. A RHEED pattern taken \textit{in situ} immediately after BN deposition reveals well-defined stripes (see inset of Fig.~\ref{TEM-LEED}(a) and Fig.\,S1 of the Supplementary material). No signatures of amorphous (diffuse-like pattern) or multicrystalline (ring-shaped pattern) phases are observed. The elongated shape of the stripes could be attributed to a rough surface, which is typical for commercially available single crystals of metals. This pattern allows a preliminary conclusion that the epitaxial growth of BN takes place and that the surface morphology resembles  that of single crystal Ni(111). High-resolution cross-sectional TEM images are presented in panels (b) and (c). These images were collected in the $<110>$ direction of bulk Ni. One can clearly see that the epitaxial growth of $h$-BN on Ni(111) is obtained with the A-B stacking of single BN layers [see inset of Fig.1(c)]. The TEM-measured interlayer distance in $h$-BN is $3.86\pm0.32$\,\AA\, while the $h$-BN-Ni separation equals $1.87\pm0.12$\,\AA. The corresponding extracted in-plane lattice constant of BN is $2.44\pm0.2$\,\AA. These values are close to the respective parameters of bulk $h$-BN: $2.504$\,\AA\ for in-plane lattice constant and $3.33$\,\AA\ for the distance between layers~\cite{Lynch:1966fk}. After synthesis, the ml-$h$-BN/Ni(111) sample was transferred under atmospheric conditions to other experimental stations for surface-sensitive experiments. The high inertness of BN to the ambient environment allowed to avoid irreversible oxidation and contamination of the sample surface ~\cite{Liu:1eg}.

Fig.~\ref{TEM-LEED}(d) shows a LEED image of the ml-$h$-BN/Ni(111) system obtained after atmospheric transfer of the sample into a UHV chamber and the respective annealing step at $400^\circ$\,C in order to remove possible air-adsorbates (water and CO). The hexagonal LEED pattern measured at the primary electron beam energy of $74$\,eV indicates the in-plane long-range order of the studied sample. 
\newline
\newline
\textbf{Electronic structure of ml-$h$-BN/Ni(111).}

\begin{figure}
\includegraphics[scale=0.15]{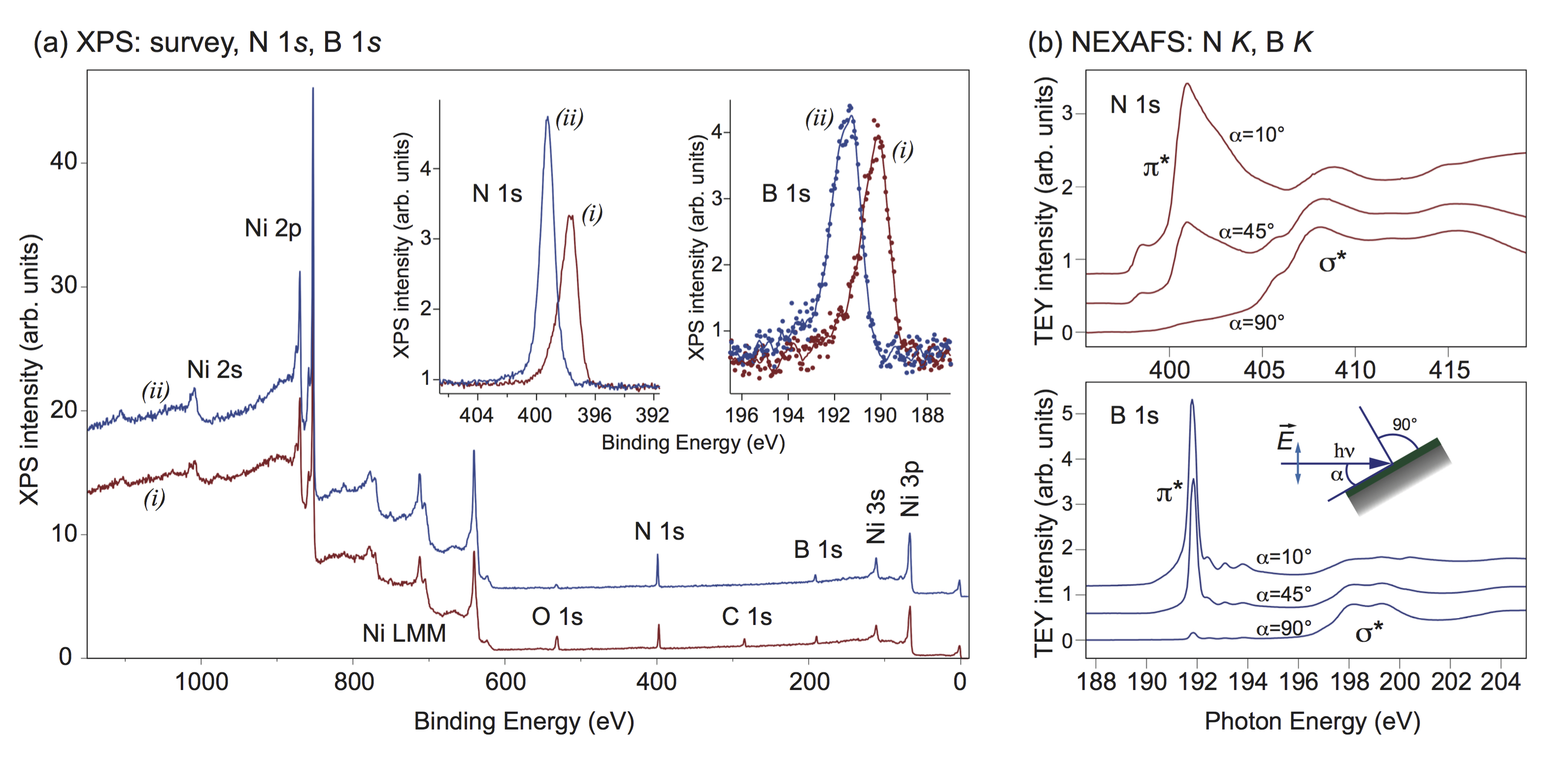}\\
\caption{\label{XPS-NEXAFS} \textbf{XPS and NEXAFS of h-BN/Ni(111).} (a) XPS survey spectra obtained with Al\,$K\alpha$ radiation for as-transferred into UHV $h$-BN/Ni(111) sample (i) and after annealing the same sample at $400^\circ$\,C (ii). The main core-level and Auger emission lines are marked in the Figure. Insets show the corresponding highly resolved XPS spectra of the N\,$1s$ and B\,$1s$ core levels. (b) TEY NEXAFS spectra of $h$-BN/Ni(111) collected as functions of photon energy at the $K$ absorption edges of N and B for different incident angles $\alpha$ between the sample plane and the incoming light (angles are marked in the figure).} 
\end{figure}

The chemical composition of the sample was verified by x-ray photoelectron spectroscopy (XPS). The results are shown in Fig.~\ref{XPS-NEXAFS}(a). This figure shows XPS spectra (survey spectra and insets with detailed N\,$1s$ and B\,$1s$ spectra) recorded (i) after the sample was transferred to the UHV chamber (brown lines) and (ii) after it was annealed in UHV at $400^\circ$\,C for $30$\,min (blue lines). This simple temperature treatment leads to the complete disappearance of the C $1s$ peak and thus to desorption of the carbon-containing molecules (CO, CO$_2$) from the surface of $h$-BN, and the XPS signal originating from oxygen is largely decreased.

The positions of the N\,$1s$ and B\,$1s$ XPS emission lines of the clean surface area are $399.25$\,eV and $191.35$\,eV, respectively, and these values are in a very good agreement with previously published data for the thick $h$-BN layer formed on polycrystalline surfaces of Cu and Ni~\cite{Liu:1eg,Kidambi:2014hv}. The analysis of the $h$-BN film composition on the basis of the corresponding XPS line intensities gives a stoichiometric compound within the error of the XPS measurements ($\pm1\%$). The subsequent angular-resolved XPS (AR-XPS) measurements of the intensity of Ni\,$2p_{3/2}$ and N\,$1s$ emission lines has been carried out (see Fig.\,S2 of Supplementary material). In this measurement the probed thickness of the BN film is varied as a function of the angle $\theta$ between the sample plane and the direction to the energy analyzer. As a result the AR-XPS method gives a thickness of the studied $h$-BN film of $40\pm5$\,\AA, which is in a good agreement with TEM data ($38.6\pm3$\,\AA) in Fig.~\ref{TEM-LEED}.

Further confirmation of the high crystalline quality of the obtained $h$-BN is revealed by NEXAFS experiments. The NEXAFS spectra were acquired in the total electron yield mode (TEY, sample drain current) for the annealed samples at the $K$ absorption edges of N and B as a function of angle $\alpha$ between sample plane and the direction of the incoming light (Fig.~\ref{XPS-NEXAFS}(b), bottom). These spectra demonstrate an example of the so-called \textit{search-light-like} effect~\cite{Stohr:1999b}, when the absorption intensity depends on the relative orientation of the electric field vector of the incoming light and the alignment of the orbital, onto which the electron is transferred from the respective core level. Therefore, the NEXAFS peaks at $398-405$\,eV for N\,$K$ and $190-194$\,eV for B\,$K$ are assigned to the transitions of $1s$ core electrons of corresponding atoms into the antibonding $\pi^*$ band of $h$-BN~\cite{Laskowski:2009}. On the other hand, the high energy tails of both NEXAFS spectra are assigned to the transitions of $1s$ electrons into the $\sigma^*$ bands of $h$-BN~\cite{Laskowski:2009}. The presented spectra are very similar to the ones of bulk $h$-BN~\cite{Preobrajenski:2004gw}. These signatures underscore the ordered crystalline structure of our epitaxial $h$-BN on Ni(111) having the electronic structure similar to that of bulk $h$-BN. The pre-edge structure in the NEXAFS spectra for both absorption edges originates from the interface states between $h$-BN and Ni(111) as a result of the orbital mixing between Ni\,$3d$ and BN\,$\pi$ states~\cite{Preobrajenski:2004gw}.

\begin{figure}
\includegraphics[scale=0.35]{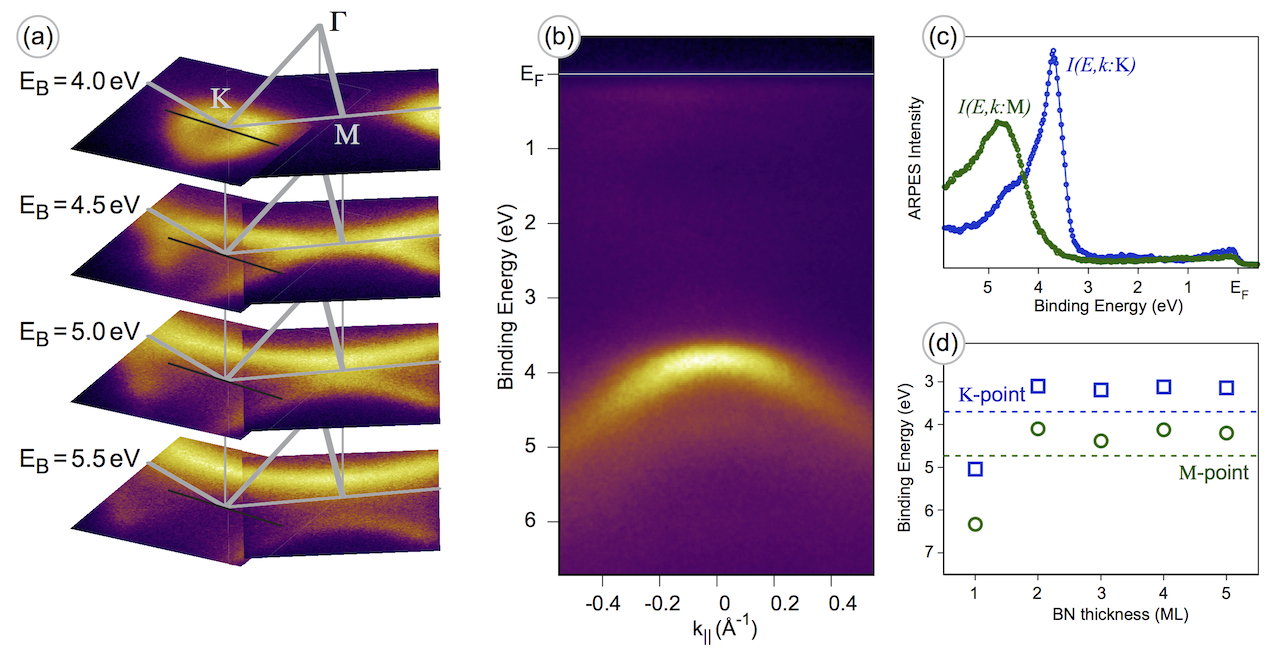}\\
\caption{\label{ARPES} \textbf{ARPES and DFT for $h$-BN/Ni(111).} (a) Series of the ARPES constant binding energy cuts for ml-$h$-BN/Ni(111). Data are extracted from the two combined data sets consecutively acquired via polar angle scanning of the sample along the $\Gamma-\mathrm{K}$ and $\Gamma-\mathrm{M}$ directions of the hexagonal BZ. The corresponding binding energies ($E_B$) for the energy cuts and the high symmetry points of the BZ are marked in the figure. (b) ARPES intensity map extracted from the 3D data set along the black line in the $k$-space as marked in (a). (c) PES intensity profiles corresponding to the $\mathrm{K}$ and $\mathrm{M}$ points extracted from the 3D data sets. (d) Calculated binding energies of the $\pi$ band of ml-$h$-BN on Ni(111) as a function of BN-thickness at the $\mathrm{K}$ (squares) and $\mathrm{M}$ (circles) points obtained in DFT calculations. Dashed horizontal lines are experimental values from (c).}
\end{figure}

An additional investigation of the electronic structure below $E_F$ of ml-$h$-BN/Ni(111) was carried out by means of angle-resolved PES (ARPES). Figure~\ref{ARPES}(a) shows constant energy cuts obtained via combination of two data sets collected as a series of angle-resolved photoemission maps along $\Gamma-\mathrm{M}$ and $\Gamma-\mathrm{K}$ directions of the Brillouine zone (BZ) of the studied system. The emission from the $h$-BN $\pi$ states is found in the valence band. This emission band demonstrates the characteristic trigonal warping around the $\mathrm{K}$ point [Fig.~\ref{ARPES}(a)]. The corresponding emission intensity map taken as a wave vector cut along the black line passing through the $\mathrm{K}$ point, marked in the panel (a), is shown in (b) as $I(E_B, k_{||})$ intensity distribution. This map and the corresponding one around the $\mathrm{M}$ point allow to extract the respective intensity profiles shown in Fig.~\ref{ARPES}(c). These data allow to find positions of the valence band maximum for different high symmetry points of the BZ: $E_B=3.70\pm0.02$\,eV for the $\mathrm{K}$ point and $E_B=4.73\pm0.02$\,eV for the $\mathrm{M}$ point. The weak band dispersion observed at higher binding energies might be assigned to the faint photoemission intensity from the deep interface BN layer (see section Discussion). 

\section*{Discussion}

The experimental results presented above were compared with the DFT calculations of electronic structure modifications of ml-$h$-BN/Ni(111) depending on the thickness of $h$-BN (for technical details, see section Methods). Fig.~\ref{ARPES}(d) shows the evolution of the binding energy of the BN\,$\pi$ band at the $\mathrm{K}$ and $\mathrm{M}$ points with the $h$-BN thickness. Despite the difficulties in the modeling of insulating compounds by means of DFT (underestimation of the band gap~\cite{Perdew:1986ft}), a careful selection of hybrid density functionals~\cite{Seidl:1996vs,Xiao:2011cg} resulted in reasonable theoretical values of $\mathrm{K}$ and $\mathrm{M}$ point binding energies for thick $h$-BN layer ($>2$\,ML), which are very close to the experimental ones (shown by the dashed lines of the corresponding colour in Fig.~\ref{ARPES}(d)).

\begin{figure}
\includegraphics[scale=0.35]{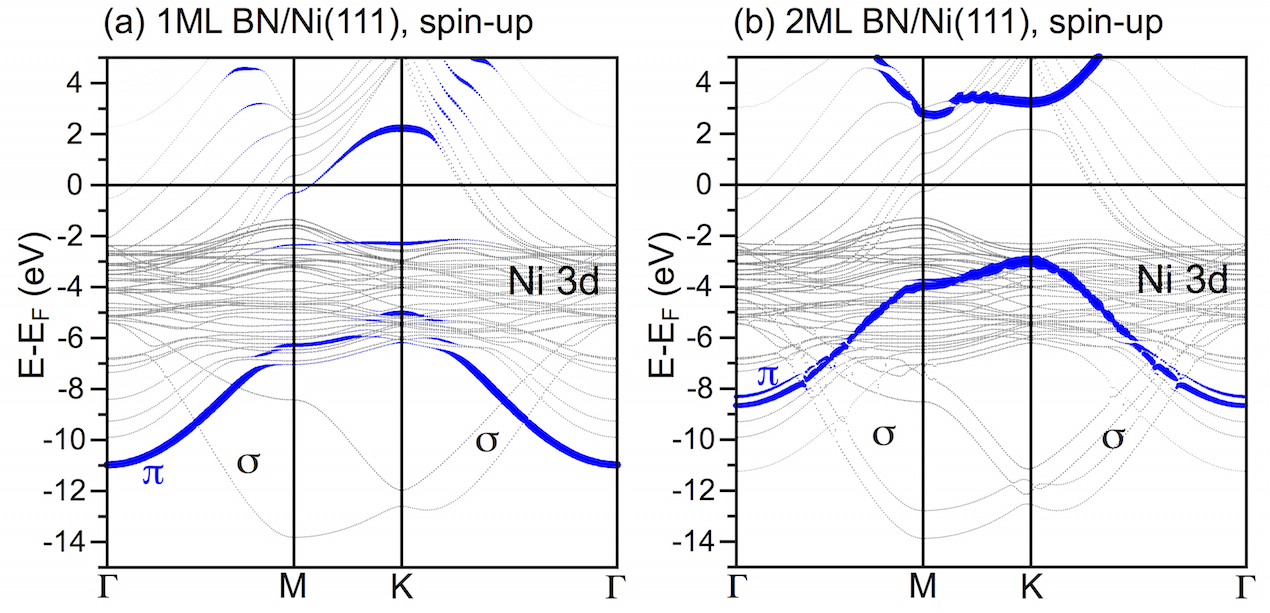}\\
\caption{\label{DFT_bands} Calculated spin-up band structures of (a) 1\,ML-thick $h$-BN and (b) 2\,ML-thick $h$-BN on Ni(111) along high-symmetry directions of the Brillouine zone (spin-up channel). Blue lines mark the weight of the $\pi$ band of the top BN layer.}
\end{figure}

A single $h$-BN layer is adsorbed on Ni(111) in the \textit{top-fcc} arrangement as deduced from our DFT calculations (see section Methods), where the N-atom of $h$-BN is placed above the Ni-atom of the interface layer, while the B-atom occupies the \textit{fcc} hollow adsorption site of Ni(111) (see Supplementary material, Fig.\,S3). The calculated mean distance between $h$-BN and Ni(111) is $2.04$\,\AA. Note: N- and B- atoms are separated from the Ni interface differently: $d\mathrm{(N-Ni-top)}=2.10$\,\AA, $d\mathrm{(B-Ni-top)}=1.98$\,\AA, the corrugation is $0.12$\,\AA. This result is consistent with the ones of the previously published works~\cite{Grad:2003gm,Laskowski:2009}. The resulting band structure of 1\,ML-$h$-BN/Ni(111) along the high symmetry directions of the hexagonal BZ is shown in Fig.~\ref{DFT_bands}(a), where it is overlaid with the weight of the BN\,$\pi$ band shown by blue lines. The binding energy of BN\,$\pi$ states is $5.04$\,eV and $6.33$\,eV for the $\mathrm{K}$ and $\mathrm{M}$ points, respectively. The interaction between the monolayer-thick $h$-BN layer and Ni(111) leads to the formation of several \textit{interface} states as a result of energy-, space-, and $k$-overlapping of the BN\,$\pi$ orbitals and Ni\,$3d$ states having $z$-components in the angular part of the wave function. This effect leads to the \textit{metallic} character of $1$\,ML-thick $h$-BN (sizeable weight of BN\,$\pi$ states at $E_F$) and \textit{pushing down} the $\pi$ bands to larger binding energies around $\mathrm{K}$ and $\mathrm{M}$ points.

In case of the A-B stacked double layer of $h$-BN on Ni(111), the second layer is completely decoupled from the underlying substrate, that leads to the restoring of the insulating nature of $h$-BN (Fig.~\ref{DFT_bands}(b)). Increasing the number of h-BN layers does not lead to qualitative changes in the DFT-obtained band structure. The A-B stacking of $h$-BN is found in our high-resolution TEM measurements [see Fig.~\ref{TEM-LEED}(b)] and is confirmed in the respective DFT calculations as the most stable configuration with the distance between BN layers of $3.07$\,\AA. The binding energy of BN\,$\pi$ states is $3.1$\,eV and $4.1$\,eV for the $\mathrm{K}$ and $\mathrm{M}$ points, respectively. For these points the calculated band gap for 2\,ML-$h$-BN/Ni(111) is $6.26$\,eV and $6.79$\,eV. These values are very close to the experimentally measured band gap for $h$-BN~\cite{Tarrio:1989ty,Kubota:2007jl,Kobayashi:2010cg}. It is interesting to note that while the structure and binding properties (like e.\,g. interlayer interaction energy) of $h$-BN as well as hybrid materials on the basis of $h$-BN can get reliable description nowadays when employing standard DFT functionals augmented by the dispersion correction (see e.g. Refs.\cite{Bucko:2010kl,Bucko:2013tl}), the reported DFT band gaps are often underestimated with respect to the experiment (e.\,g. Ref.\cite{Perdew:1996} predict $E_g=4.56$\,eV for the $h$-BN sheet~\cite{Berseneva:2013bf}). The Heyd-Scuseria-Ernzerhof (HSE) functional used in the present study predicts the $h$-BN gap very close to experiment~\cite{Berseneva:2013bf,Park:2012kd}. 

\section*{Conclusion}

We have shown that the epitaxial growth of an $h$-BN multilayer (ml-$h$-BN) is feasible on the close-packed Ni(111) surface of a Ni single crystal by means of MBE. The obtained epitaxial films of ml-$h$-BN demonstrate very high structural perfection with A-B stacking of BN layers as observed using cross-sectional TEM. Photoelectron spectroscopy experiments demonstrate clearly the energy dispersion of the valence band states below the Fermi level. These data in combination with results of NEXAFS and DFT calculations allow for the systematic mapping of the electronic structure of the system. The DFT calculations show good agreement with the experimental data and with bulk electronic properties of $h$-BN. DFT also reveals that the first $h$-BN layer at the interface of Ni/ml-$h$-BN is metallic, thus placing the metal/insulator interface inside of ml-$h$-BN. Our findings demonstrate that MBE growth of ml-$h$-BN is a perspective method that can be easily adapted to electronic and spintronic applications of $h$-BN. We propose that this epitaxial Ni(111)/ml-$h$-BN system, can be used as an effective substrate for the fabrication of FM/BN/graphene spin injection devices or FM/BN/graphene/BN/FM spin switches (FM=ferromagnet), where graphene is effectively decoupled from the metallic substrate. This allows to preserve the unique band structure of graphene in the vicinity of $E_F$.

\section*{Methods}

\textbf{Samples preparation.} The growth of $h$-BN films was carried out by applying molecular beam epitaxy (MBE, Riber SIVA 45 setup) on a Ni(111) surface. The growth recipe was combined out of two growth approaches described elsewhere~\cite{Tsai:2009jw,Nakhaie:2015vl}. A single crystalline Ni(111) was used as a substrate. The cleaning of the substrates consisted of a heating step to the temperature of approximately $800^\circ$\,C followed by the deposition of a $20$\,nm thick Ni buffer layer using e-beam evaporation at the substrate temperature of $700^\circ$\,C. (No sputtering was used during cleaning of Ni(111) substrates.) Then, an equivalent of one monolayer (ML) amount of B was deposited on the substrate followed by the deposition of BN at $700^\circ$\,C. The flux of boron atoms was generated by using a high-temperature effusion cell operated at $1900^\circ$\,C. A preliminary calibration of the B-flux was carried out by applying secondary ion mass spectroscopy on a uniformly doped Si:B layer grown on a Si(001) substrate. The flux of nitrogen was obtained by an RF-plasma source (Riber RF-N 50-63) at the RF-power of $250$\,W and the MBE background pressure of $1.3\times10^{-5}$\,Torr. The growth rate of BN was $0.14$\,ML/min. 

\textbf{Sample characterization.} The as-grown $h$-BN films were characterized \textit{in situ} by using reflection high energy electron diffraction (RHEED) and Auger electron spectroscopy (AES). The \textit{ex situ} characterization was carried out outside the MBE system requiring sample transfer through the atmosphere and, in some cases, additional preparations. The morphology and crystal structure of the films were analyzed by transmission electron microscopy (TEM) and scanning TEM (STEM). The corresponding specimen of specific sample orientation have been generated by Focused Ion Beam (FIB) preparation technique. TEM investigations were carried out in a TITAN 80/300 electron microscope. A low energy electron diffraction (LEED) was carried out in the SPECS UHV probe station. In some cases of \textit{ex situ} characterization (LEED, XPS, ARPES) the samples were annealed at $400^\circ$\,C. A set of surface-sensitive characterization methods of the $h$-BN electronic structure is carried out and described below.

\textbf{NEXAFS experiments.} These experiments were performed at the Russian-German beamline of the BESSY\,II storage ring of HZB, Berlin. X-ray absorption spectroscopy experiments were performed in the total electron yield (TEY) mode via measurement of the drain current. Spectra were collected around $K$ absorption edges of N and B with the total resolution of the beamline of $50$\,meV. Absorption experiments performed at different angles between sample surface and the direction of the incoming light allow to verify spacial orientation of the valence band states via the so-called \textit{search-light-like} effect.

\textbf{ARPES experiments.} The ARPES measurements  with He\,II radiation (photon energy of $h\nu=40.81$\,eV) were performed in the SPECS demo lab using a FlexPS system with PHOIBOS 150 2D-CCD analyzer. In this case a 5-axis motorized manipulator was used, allowing for a precise alignment of the sample in $k$ space. The sample was azimuthally pre-aligned in such a way that the polar scans were performed along the $\Gamma-\mathrm{K}$ or $\Gamma-\mathrm{M}$ directions of the hexagonal Brillouine zone of $h$-BN/Ni(111) with the photoemission intensity on the channelplate images acquired along the direction perpendicular to $\Gamma-\mathrm{K}$ or $\Gamma-\mathrm{M}$ directions, respectively. The final 3D data sets of the photoemission intensity as a function of kinetic energy and two emission angles, $I(E_{kin},angle_1,angle_2)$, were then carefully analyzed.

\textbf{DFT calculations.} Calculations were performed using the plane-wave projector augmented-wave (PAW)~\cite{Blochl:1994,Kresse:1999} method applying the semi-local Perdew-Burke-Ernzerhof (PBE)~\cite{Perdew:1996} exchange-correlation functional augmented by the DFT-D2 method of Grimme~\cite{Grimme:2006} for describing the dispersion interactions and the Heyd-Scuseria-Ernzerhof (HSE)~\cite{Heyd:2003eg} hybrid functional as implemented in the Vienna \textit{ab initio} simulation package (VASP)~\cite{Kresse:1996a,Kresse:1996} code. The HSE screening parameter was set to a value of $0.2\,\mathrm{\AA}^{-1}$~\cite{Krukau:2006jq}. The plane-wave cutoff energy was set to $500$\,eV. Brillouin-zone integration was performed on $\Gamma$-centered symmetry reduced Monkhorst-Pack~\cite{Monkhorst:1976} meshes using a Methfessel-Paxton smearing method of first order with $\sigma=0.2$\,eV, except for the calculation of total energies and densities of states (DOSs). For those calculations, the tetrahedron method with Bl\"ochl corrections~\cite{Blochl:1994vg} was used. A $12\times 12\times 1$ k-mesh was used. Although for both computational schemes (HSE and PBE) the qualitative results are found to be similar, the HSE functional predicts the band gaps reasonably well, while the PBE functional yields a significant underestimation of the band gaps. ml-$h$-BN/Ni(111) was modelled, in case of relaxation (band structure) calculations, as a stack of 13\,ML (7\,ML) separated by a spacer vacuum of $45$\,\AA\ with the corresponding number of $h$-BN layers in A-B stacking on both (one) sides of the Ni-slab.


\begin{thebibliography}{10}

\bibitem{Novoselov:2004a}
K.~Novoselov et~al.,
\newblock Science {\bf 306}, 666 (2004).

\bibitem{Novoselov:2005}
K.~Novoselov et~al.,
\newblock Nature {\bf 438}, 197 (2005).

\bibitem{Zhang:2005}
Y.~Zhang, Y.~Tan, H.~Stormer, and P.~Kim,
\newblock Nature {\bf 438}, 201 (2005).

\bibitem{Tombros:2007}
N.~Tombros, C.~Jozsa, M.~Popinciuc, H.~T. Jonkman, and B.~J. van Wees,
\newblock Nature {\bf 448}, 571 (2007).

\bibitem{Yang:2011}
T.-Y. Yang et~al.,
\newblock Phys. Rev. Lett. {\bf 107}, 047206 (2011).

\bibitem{Han:2011id}
W.~Han and R.~K. Kawakami,
\newblock Phys. Rev. Lett. {\bf 107}, 047207 (2011).

\bibitem{Bader:2010gr}
S.~D. Bader and S.~S.~P. Parkin,
\newblock Annu. Rev. Condens. Matter Phys. {\bf 1}, 71 (2010).

\bibitem{Geim:2007a}
A.~K. Geim and K.~S. Novoselov,
\newblock Nature Materials {\bf 6}, 183 (2007).

\bibitem{CastroNeto:2009}
A.~Castro~Neto, F.~Guinea, N.~Peres, K.~Novoselov, and A.~Geim,
\newblock Rev. Mod. Phys. {\bf 81}, 109 (2009).

\bibitem{Novoselov:2011}
K.~Novoselov,
\newblock Rev. Mod. Phys. {\bf 83}, 837 (2011).

\bibitem{Amet:2015ka}
F.~Amet et~al.,
\newblock Nature Communications {\bf 6}, 5838 (2015).

\bibitem{Faugeras:2015it}
C.~Faugeras et~al.,
\newblock Phys. Rev. Lett. {\bf 114}, 126804 (2015).

\bibitem{Natterer:2015kv}
F.~D. Natterer et~al.,
\newblock Phys. Rev. Lett. {\bf 114}, 245502 (2015).

\bibitem{Kumar:2015ct}
A.~Kumar, T.~Low, K.~H. Fung, P.~Avouris, and N.~X. Fang,
\newblock Nano Lett. {\bf 15}, 3172 (2015).

\bibitem{Gao:2015jl}
Y.~Gao et~al.,
\newblock Nano Lett. {\bf 15}, 2001 (2015).

\bibitem{Gaskell:2015io}
J.~Gaskell et~al.,
\newblock Appl. Phys. Lett. {\bf 107}, 103105 (2015).

\bibitem{Parzefall:2015eo}
M.~Parzefall et~al.,
\newblock Nature Nanotechnology , 1 (2015).

\bibitem{Petrone:2015kk}
N.~Petrone et~al.,
\newblock ACS Nano {\bf 9}, 8953 (2015).

\bibitem{Banszerus:2015bx}
L.~Banszerus et~al.,
\newblock Science Advances {\bf 1}, e1500222 (2015).

\bibitem{Drogeler:2014cw}
M.~Dr{\"o}geler et~al.,
\newblock Nano Lett. {\bf 14}, 6050 (2014).

\bibitem{Kamalakar:2015is}
M.~V. Kamalakar, C.~Groenveld, A.~e. Dankert, and S.~P. Dash,
\newblock Nature Communications {\bf 6}, 6766 (2015).

\bibitem{Tarrio:1989ty}
C.~Tarrio and S.~Schnatterly,
\newblock Phys. Rev., B Condens. Matter {\bf 40}, 7852 (1989).

\bibitem{Kubota:2007jl}
Y.~Kubota, K.~Watanabe, O.~Tsuda, and T.~Taniguchi,
\newblock Science {\bf 317}, 932 (2007).

\bibitem{Lee:2011cf}
G.-H. Lee et~al.,
\newblock Appl. Phys. Lett. {\bf 99}, 243114 (2011).

\bibitem{Britnell:2013ca}
L.~Britnell et~al.,
\newblock Science {\bf 340}, 1311 (2013).

\bibitem{Kirczenow:2001ef}
G.~Kirczenow,
\newblock Phys. Rev. B {\bf 63}, 054422 (2001).

\bibitem{Parkin:2004do}
S.~S.~P. Parkin et~al.,
\newblock Nature Materials {\bf 3}, 862 (2004).

\bibitem{Dankert:2014ip}
A.~Dankert, M.~Venkata~Kamalakar, A.~Wajid, R.~S. Patel, and S.~P. Dash,
\newblock Nano Res. {\bf 8}, 1357 (2014).

\bibitem{Yazyev:2009}
O.~V. Yazyev and A.~Pasquarello,
\newblock Phys. Rev. B {\bf 80}, 035408 (2009).

\bibitem{Ismach:2012hn}
A.~Ismach et~al.,
\newblock ACS Nano {\bf 6}, 6378 (2012).

\bibitem{Lu:2015ev}
G.~Lu et~al.,
\newblock Nature Communications {\bf 6}, 6160 (2015).

\bibitem{Sutter:2013eu}
P.~Sutter, J.~Lahiri, P.~Zahl, B.~Wang, and E.~Sutter,
\newblock Nano Lett. {\bf 13}, 276 (2013).

\bibitem{Nakhaie:2015vl}
S.~Nakhaie et~al.,
\newblock Appl. Phys. Lett. {\bf 106}, 213108 (2015).

\bibitem{Lynch:1966fk}
R.~W. Lynch,
\newblock J. Chem. Phys. {\bf 44}, 181 (1966).

\bibitem{Liu:1eg}
Z.~Liu et~al.,
\newblock Nature Communications {\bf 4}, 2541 (2013).

\bibitem{Kidambi:2014hv}
P.~R. Kidambi et~al.,
\newblock Chem. Mater. {\bf 26}, 6380 (2014).

\bibitem{Stohr:1999b}
J.~St{\"o}hr and M.~Samant,
\newblock J. Electr. Spectr. Rel. Phenom. {\bf 98}, 189 (1999).

\bibitem{Laskowski:2009}
R.~Laskowski, T.~Gallauner, P.~Blaha, and K.~Schwarz,
\newblock J Phys.: Condens. Matter {\bf 21}, 104210 (2009).

\bibitem{Preobrajenski:2004gw}
A.~Preobrajenski, A.~Vinogradov, and N.~Martensson,
\newblock Phys. Rev. B {\bf 70}, 165404 (2004).

\bibitem{Perdew:1986ft}
J.~P. Perdew,
\newblock Int. J. Quantum. Chem. {\bf 19}, 497 (1986).

\bibitem{Seidl:1996vs}
A.~Seidl, A.~G{\"o}rling, P.~Vogl, J.~Majewski, and M.~Levy,
\newblock Phys. Rev. B {\bf 53}, 3764 (1996).

\bibitem{Xiao:2011cg}
H.~Xiao, J.~Tahir-Kheli, and W.~A. Goddard, III,
\newblock J. Phys. Chem. Lett. {\bf 2}, 212 (2011).

\bibitem{Grad:2003gm}
G.~B. Grad, P.~Blaha, K.~Schwarz, W.~Auwarter, and T.~Greber,
\newblock Phys. Rev. B {\bf 68}, 085404 (2003).

\bibitem{Kobayashi:2010cg}
Y.~Kobayashi, C.-L. Tsai, and T.~Akasaka,
\newblock Phys. Status Solidi (c) {\bf 7}, 1906 (2010).

\bibitem{Bucko:2010kl}
T.~Bu{\v c}ko, J.~Hafner, S.~Leb{\`e}gue, and J.~G. {\'A}ngy{\'a}n,
\newblock J. Phys. Chem. A {\bf 114}, 11814 (2010).

\bibitem{Bucko:2013tl}
T.~Bu{\v c}ko, S.~Leb{\`e}gue, J.~Hafner, and J.~G. {\'A}ngy{\'a}n,
\newblock Phys. Rev. B {\bf 87}, 064110 (2013).

\bibitem{Perdew:1996}
J.~Perdew, K.~Burke, and M.~Ernzerhof,
\newblock Phys. Rev. Lett. {\bf 77}, 3865 (1996).

\bibitem{Berseneva:2013bf}
N.~Berseneva, A.~Gulans, A.~V. Krasheninnikov, and R.~M. Nieminen,
\newblock Phys. Rev. B {\bf 87}, 035404 (2013).

\bibitem{Park:2012kd}
H.~Park, A.~Wadehra, J.~W. Wilkins, and A.~H. Castro~Neto,
\newblock Appl. Phys. Lett. {\bf 100}, 253115 (2012).

\bibitem{Tsai:2009jw}
C.~L. Tsai, Y.~Kobayashi, T.~Akasaka, and M.~Kasu,
\newblock Journal of Crystal Growth {\bf 311}, 3054 (2009).

\bibitem{Blochl:1994}
P.~E. Bl{\"o}chl,
\newblock Phys. Rev. B {\bf 50}, 17953 (1994).

\bibitem{Kresse:1999}
G.~Kresse and D.~Joubert,
\newblock Phys. Rev. B {\bf 59}, 1758 (1999).

\bibitem{Grimme:2006}
S.~Grimme,
\newblock J. Comput. Chem. {\bf 27}, 1787 (2006).

\bibitem{Heyd:2003eg}
J.~Heyd, G.~E. Scuseria, and M.~Ernzerhof,
\newblock J. Chem. Phys. {\bf 118}, 8207 (2003).

\bibitem{Kresse:1996a}
G.~Kresse and J.~Furthmuller,
\newblock Comp. Mater. Sci. {\bf 6}, 15 (1996).

\bibitem{Kresse:1996}
G.~Kresse and J.~Furthmuller,
\newblock Phys. Rev. B {\bf 54}, 11169 (1996).

\bibitem{Krukau:2006jq}
A.~V. Krukau, O.~A. Vydrov, A.~F. Izmaylov, and G.~E. Scuseria,
\newblock J. Chem. Phys. {\bf 125}, 224106 (2006).

\bibitem{Monkhorst:1976}
H.~J. Monkhorst and J.~D. Pack,
\newblock Phys. Rev. B {\bf 13}, 5188 (1976).

\bibitem{Blochl:1994vg}
P.~E. Bl{\"o}chl, O.~Jepsen, and O.~Andersen,
\newblock Phys. Rev., B Condens. Matter {\bf 49}, 16223 (1994).

\end{thebibliography}

\section*{Acknowledgements}

Financial support from the German Research Foundation (DFG) through the grant VO1711/3-1 within the Priority Program 1459 is appreciated. B.S. acknowledges St. Petersburg State University for the financial support in the framework of the grant 11.37.634.2013. The High Performance Computing Network of Northern Germany (HLRN) is acknowledged for computer time. E.N.V. would like to thank Dr. J. Paier (HU Berlin) for useful discussions. A.T. acknowledges valuable discussions with J.M.J. Lopes (PDI Berlin). M.Krause (Fraunhofer IWM Halle) is acknowledged for preliminary SIMS investigations of Si:B structures.

\section*{Author contributions}

A.T. carried out sample preparation and initial \textit{in situ} characterization. H.B. and P.W. carried out the FIB sample preparation and high-resolution TEM measurements. B.S. and Yu.S.D. performed LEED, NEXAFS, and core-level PES experiments at BESSY. E.N.V. performed DFT modelling. Yu.S.D. carried out XPS and ARPES experiments in the lab. A.T., H.B., P.W., G.G., B.S., E.N.V., S.S.P.P., and Yu.S.D. contribute in the analysis of data and discussions. Yu.S.D. wrote manuscript with contributions from all co-authors. 

\section*{Additional information}

\textbf{Competing financial interests:} The authors declare no competing financial interests.

\newpage

\newpage

\newpage

\newpage

\newpage

\noindent
Supplementary material for manuscript:\\
\textbf{Structural and electronic properties of epitaxial multilayer \textit{h}-BN on Ni(111) for spintronics applications}\\
\newline
A. A. Tonkikh,$^{1,2}$ E. N. Voloshina,$^3$ P. Werner,$^1$ H. Blumtritt,$^1$  B. Senkovskiy,$^{4,5}$ G. G\"untherodt,$^{1,6}$ S. S. P. Parkin,$^1$ and Yu. S. Dedkov$^{7}$\\
\newline
$^1$Max Planck Institute of Microstructure Physics, Weinberg 2, 06120 Halle (Saale), Germany\\
$^2$Institute for Physics of Microstructures RAS , 603950, GSP-105, Nizhny Novgorod, Russia\\
$^3$Humboldt-Universit\"at zu Berlin, Institut f\"ur Chemie, 10099 Berlin, Germany\\
$^4$Institute of Solid State Physics, Dresden University of Technology, 01062 Dresden, Germany\\
$^5$St. Petersburg State University, 198504 St. Petersburg, Russia\\
$^6$2nd Institute of Physics and JARA-FIT, RWTH Aachen University, 52074 Aachen, Germany\\
$^7$SPECS Surface Nano Analysis GmbH, Voltastra\ss e 5, 13355 Berlin, Germany
\newline
\textbf{List of tables and figures:}
\newline
\noindent\textbf{Fig.\,S1.} RHEED image of a 10\,ML-thick $h$-BN film on Ni(111).
\newline
\noindent\textbf{Fig.\,S2.} Angular resolved XPS spectra (Ni\,$2p_{3/2}$ and N\,$1s$) at different take-off angles $\theta$.
\newline
\noindent\textbf{Table I} Mean interlayer distances in the ml-h-BN/Ni(111) system for different thickness of the BN layer
\newline
\noindent\textbf{Fig.\,S3.} Structure of the ml-h-BN/Ni(111) system where h-BN has an A-B stacking. Distances used in Table\,I are marked in the figure.
\newline
\noindent\textbf{Fig.\,S4.} Layer resolved partial density of states (PDOS) for $1$\,ML-h-BN/Ni(111) (spin-up -- top panels; spin-down -- bottom panels).
\newline
\noindent\textbf{Fig.\,S5.} Layer resolved partial density of states (PDOS) for $2$\,ML-h-BN/Ni(111) (spin-up -- top panels; spin-down -- bottom panels).
\newline
\noindent\textbf{Fig.\,S6.} Layer resolved partial density of states (PDOS) for $3$\,ML-h-BN/Ni(111) (spin-up -- top panels; spin-down -- bottom panels).
\newline
\noindent\textbf{Fig.\,S7.} Layer resolved partial density of states (PDOS) for $4$\,ML-h-BN/Ni(111) (spin-up -- top panels; spin-down -- bottom panels).
\newline
\noindent\textbf{Fig.\,S8.} Layer resolved partial density of states (PDOS) for $5$\,ML-h-BN/Ni(111) (spin-up -- top panels; spin-down -- bottom panels).
\newline

\clearpage

\begin{figure}
\includegraphics[width=0.7\textwidth]{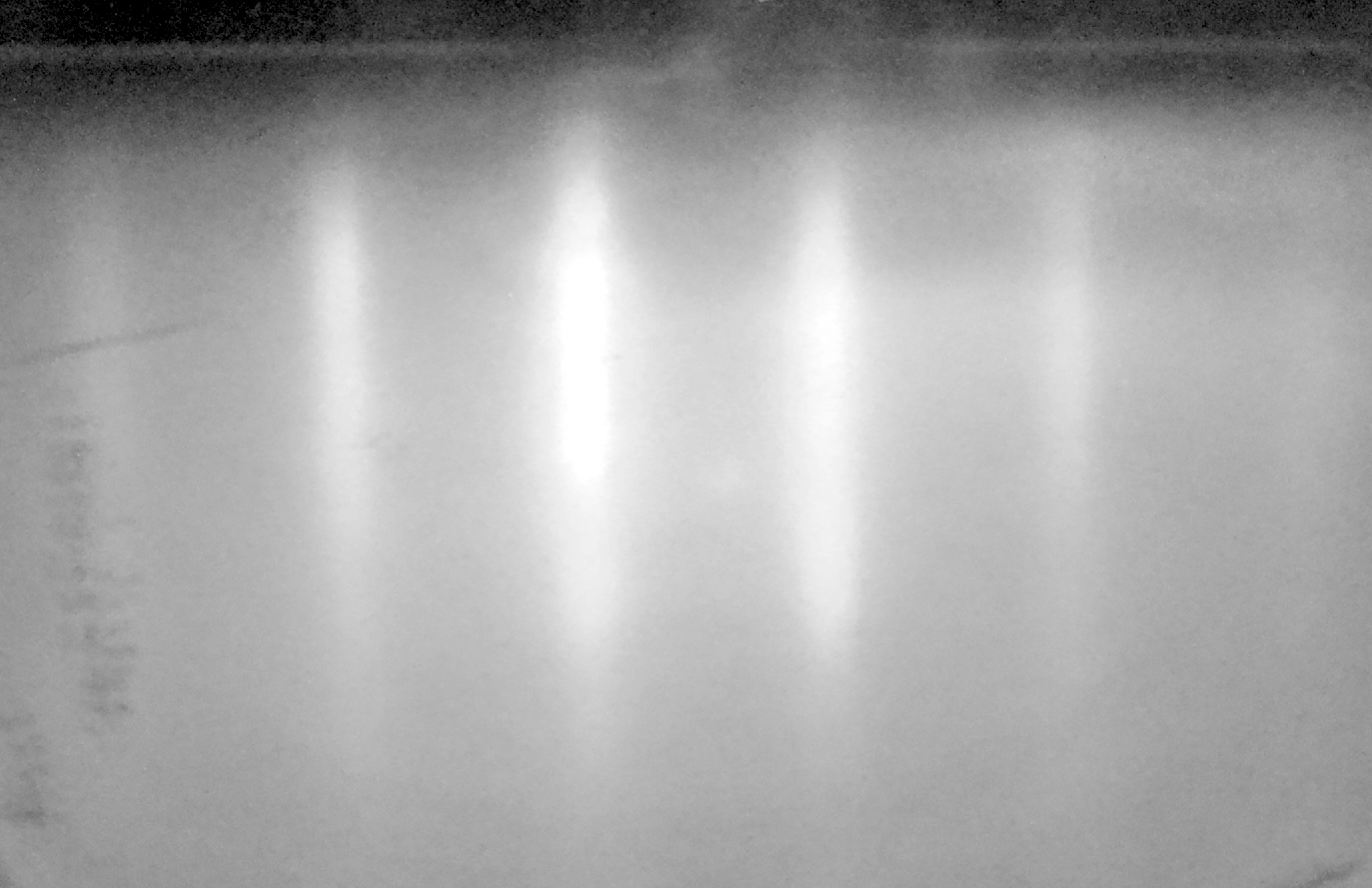}
\end{figure}
\noindent\textbf{Fig.\,S1.} RHEED image of a 10\,ML-thick $h$-BN film on Ni(111).

\clearpage

\begin{figure}
\includegraphics[width=1\textwidth]{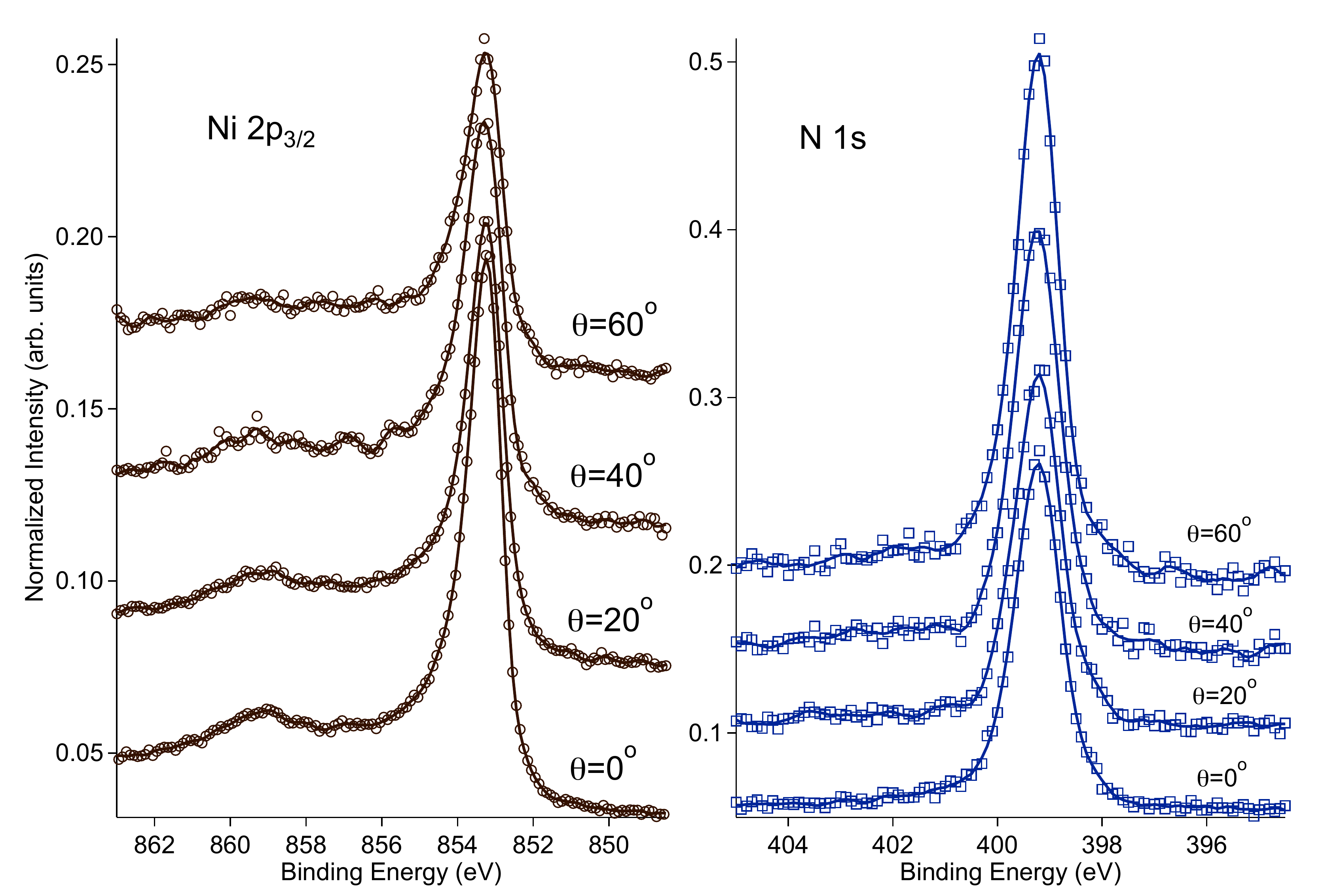}
\end{figure}
\noindent\textbf{Fig.\,S2.} Angular resolved XPS spectra (Ni\,$2p_{3/2}$ and N\,$1s$) at different take-off angles $\theta$.

\clearpage

\begin{table}
\caption{Mean interlayer distances in the ml-h-BN/Ni(111) system for different thickness of the BN layer (see Fig.\,S1).}
\begin{tabular}{p{1.5cm} p{2.8cm}p{2.8cm}p{2.8cm}p{2.8cm}p{2.8cm}}
\hline
		&1ML BN/Ni	&2ML BN/Ni	&3ML BN/Ni	&4ML BN/Ni	&5ML BN/Ni	\\
\hline
$d_1$	&$2.04$			&$2.04$			&$2.04$			&$2.04$			&$2.04$\\
$d_2$	&$2.02$			&$2.02$			&$2.02$			&$2.02$			&$2.02$\\
$d_3$	&$2.00$			&$2.00$			&$2.00$			&$2.00$			&$2.00$\\
$d_4$	&$2.04$			&$2.03$			&$2.03$			&$2.03$			&$2.03$\\
$d_5$	&				&$3.01$			&$3.00$			&$3.00$			&$3.00$\\
$d_6$	&				&				&$3.06$			&$3.05$			&$3.05$\\
$d_7$	&				&				&				&$3.06$			&$3.05$\\
$d_8$	&				&				&				&				&$3.06$\\
\hline
\\
\\
\\
\end{tabular}
\end{table}

\clearpage

\begin{figure}
\includegraphics[width=0.3\textwidth]{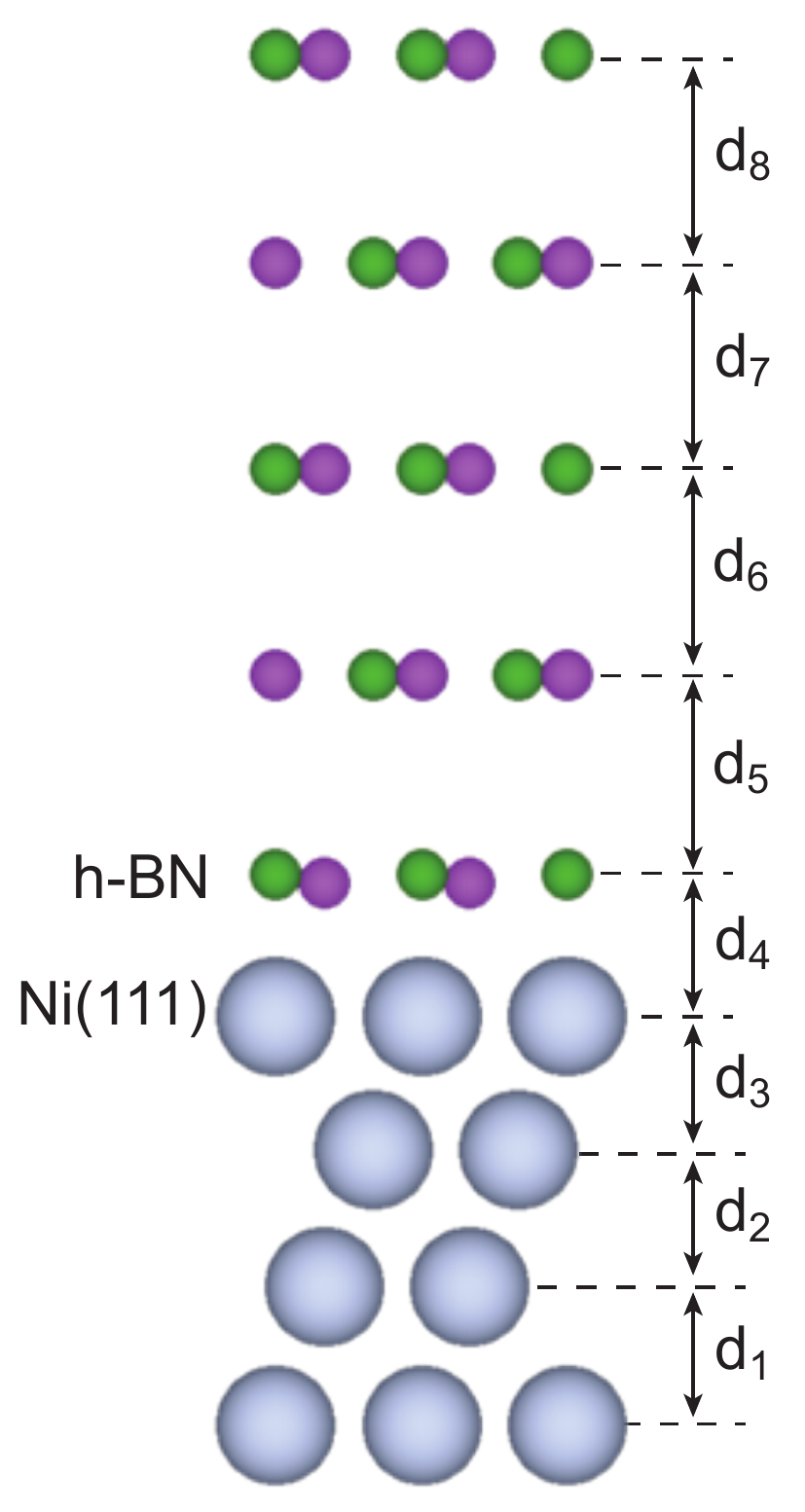}
\end{figure}
\noindent\textbf{Fig.\,S3.} Structure of the ml-h-BN/Ni(111) system where h-BN has an A-B stacking. Distances used in Table\,I are marked in the figure. 

\clearpage

\begin{figure}
\includegraphics[width=0.48\textwidth]{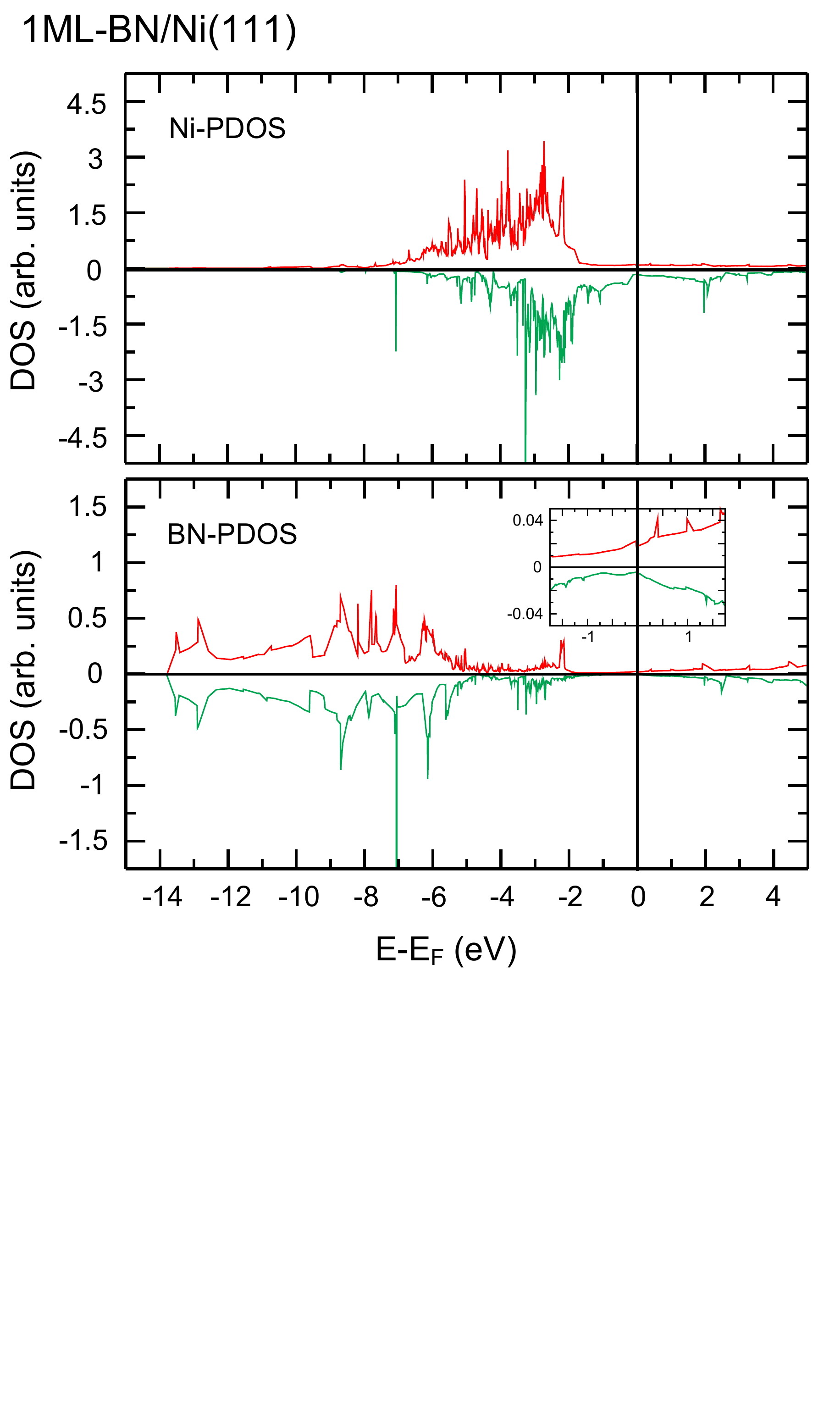}
\end{figure}
\noindent\textbf{Fig.\,S4.} Layer resolved partial density of states (PDOS) for $1$\,ML-h-BN/Ni(111) (spin-up -- top panels; spin-down -- bottom panels).

\clearpage

\begin{figure}
\includegraphics[width=0.48\textwidth]{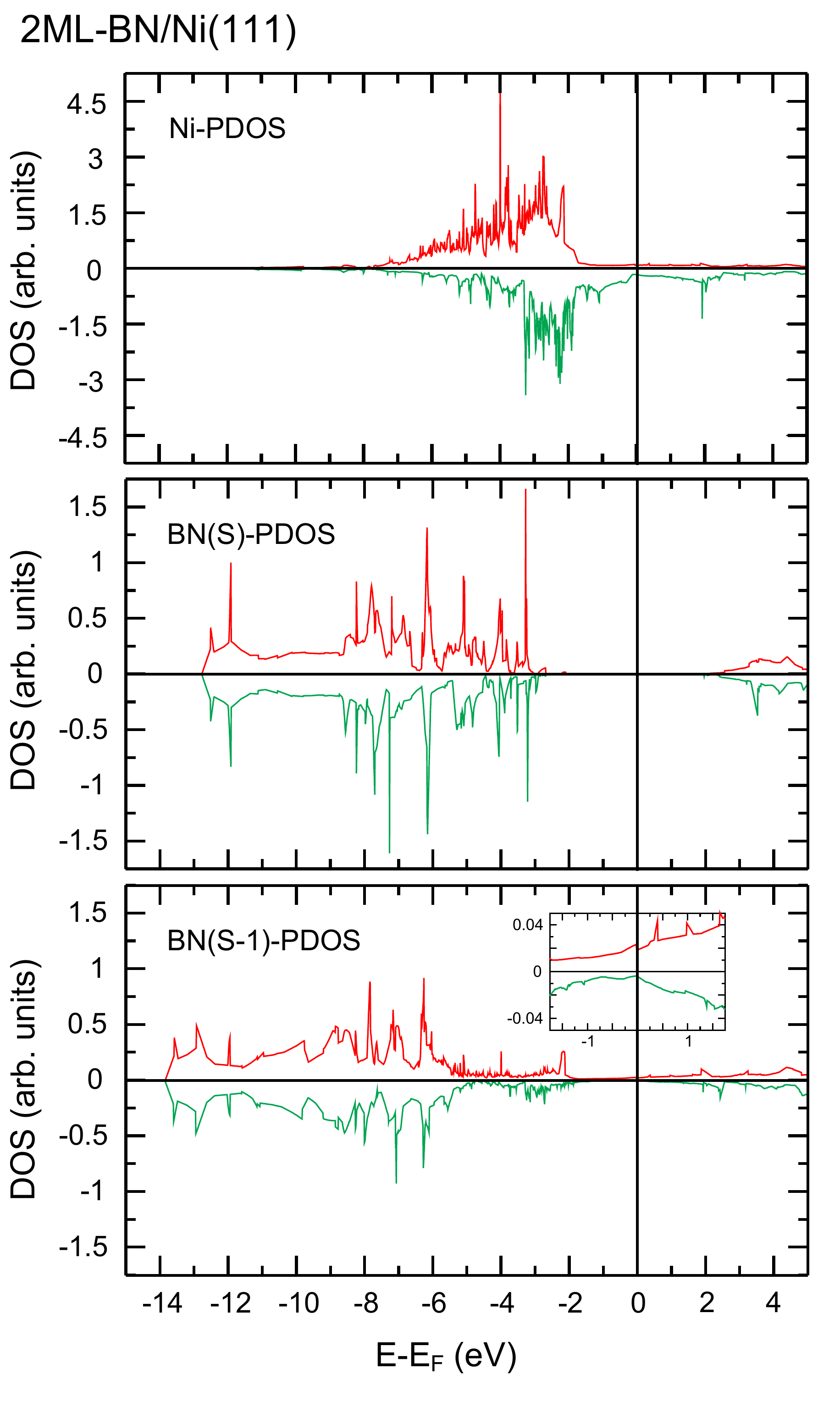}
\end{figure}
\noindent\textbf{Fig.\,S5.} Layer resolved partial density of states (PDOS) for $2$\,ML-h-BN/Ni(111) (spin-up -- top panels; spin-down -- bottom panels).

\clearpage

\begin{figure}
\includegraphics[width=0.48\textwidth]{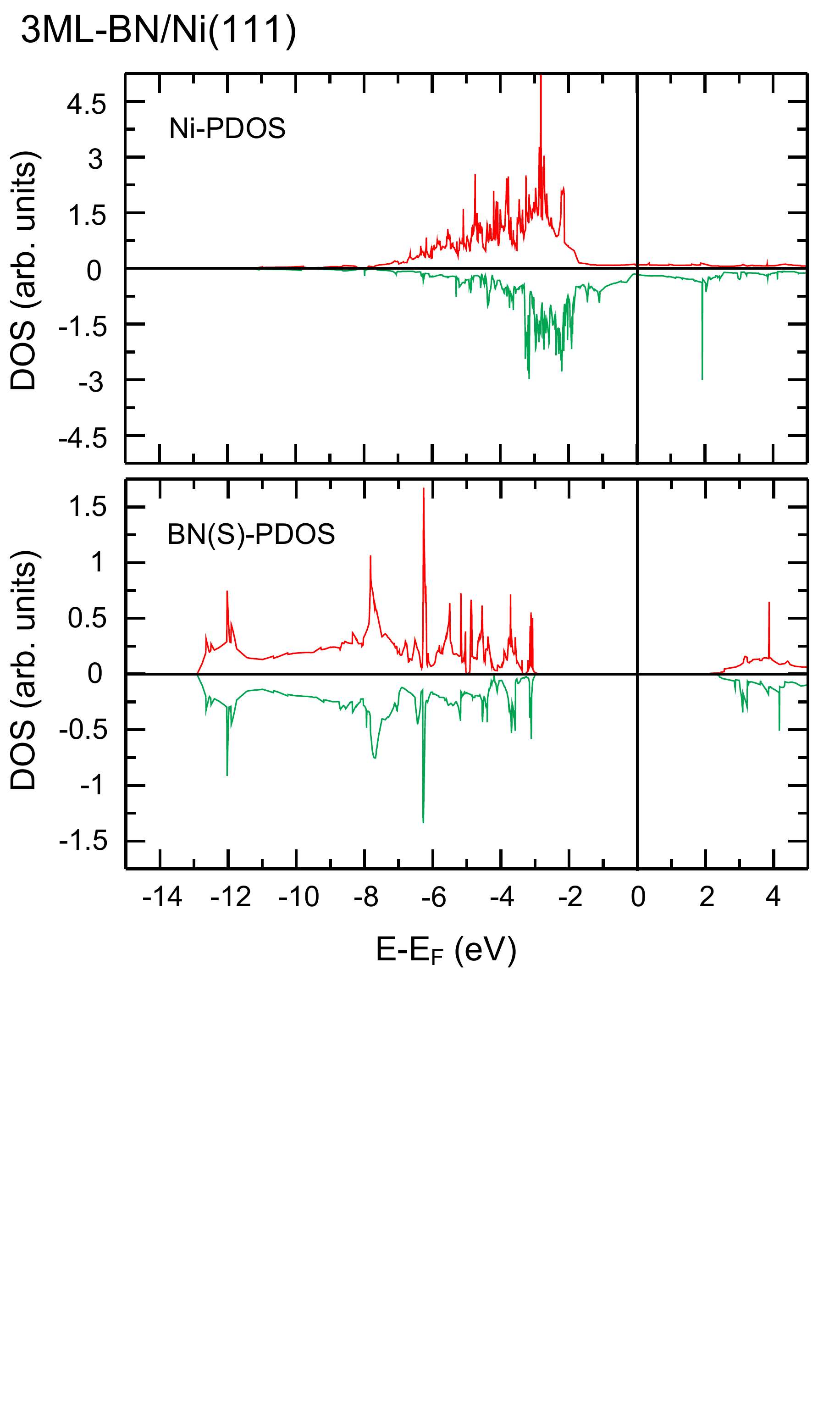}
\hspace{-0.75cm}\includegraphics[width=0.48\textwidth]{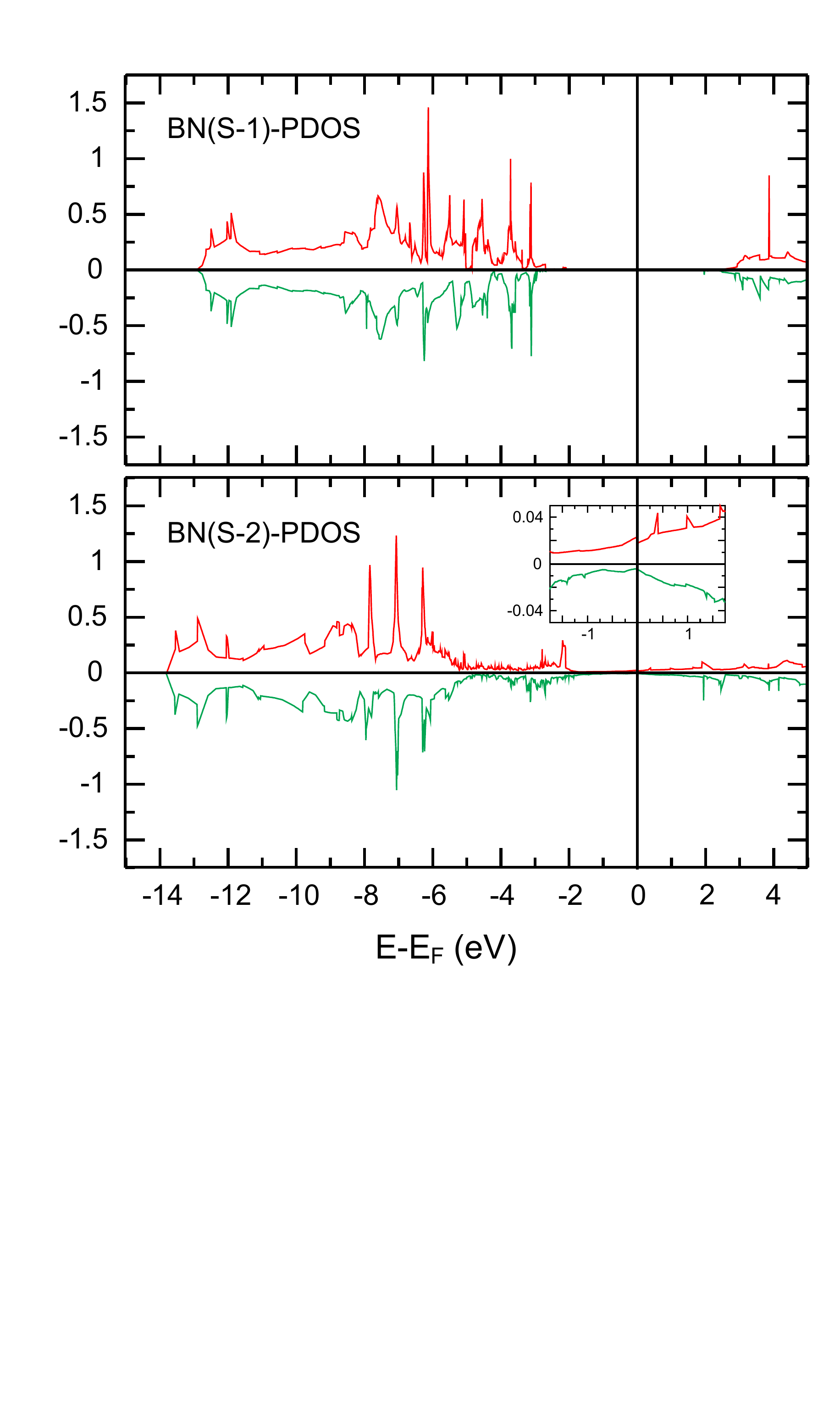}
\end{figure}
\noindent\textbf{Fig.\,S6.} Layer resolved partial density of states (PDOS) for $3$\,ML-h-BN/Ni(111) (spin-up -- top panels; spin-down -- bottom panels).

\clearpage

\begin{figure}
\includegraphics[width=0.48\textwidth]{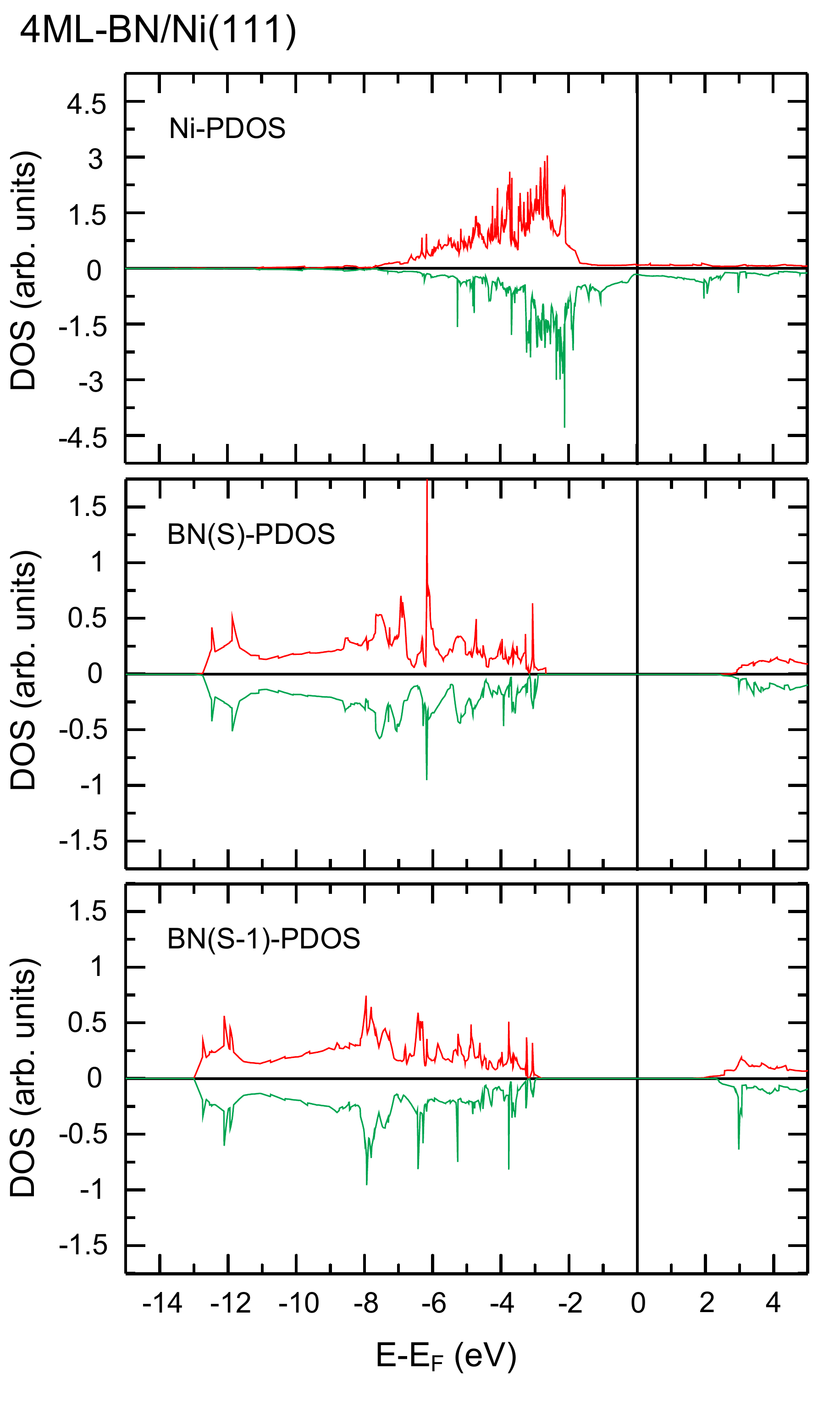}
\hspace{-0.75cm}\includegraphics[width=0.48\textwidth]{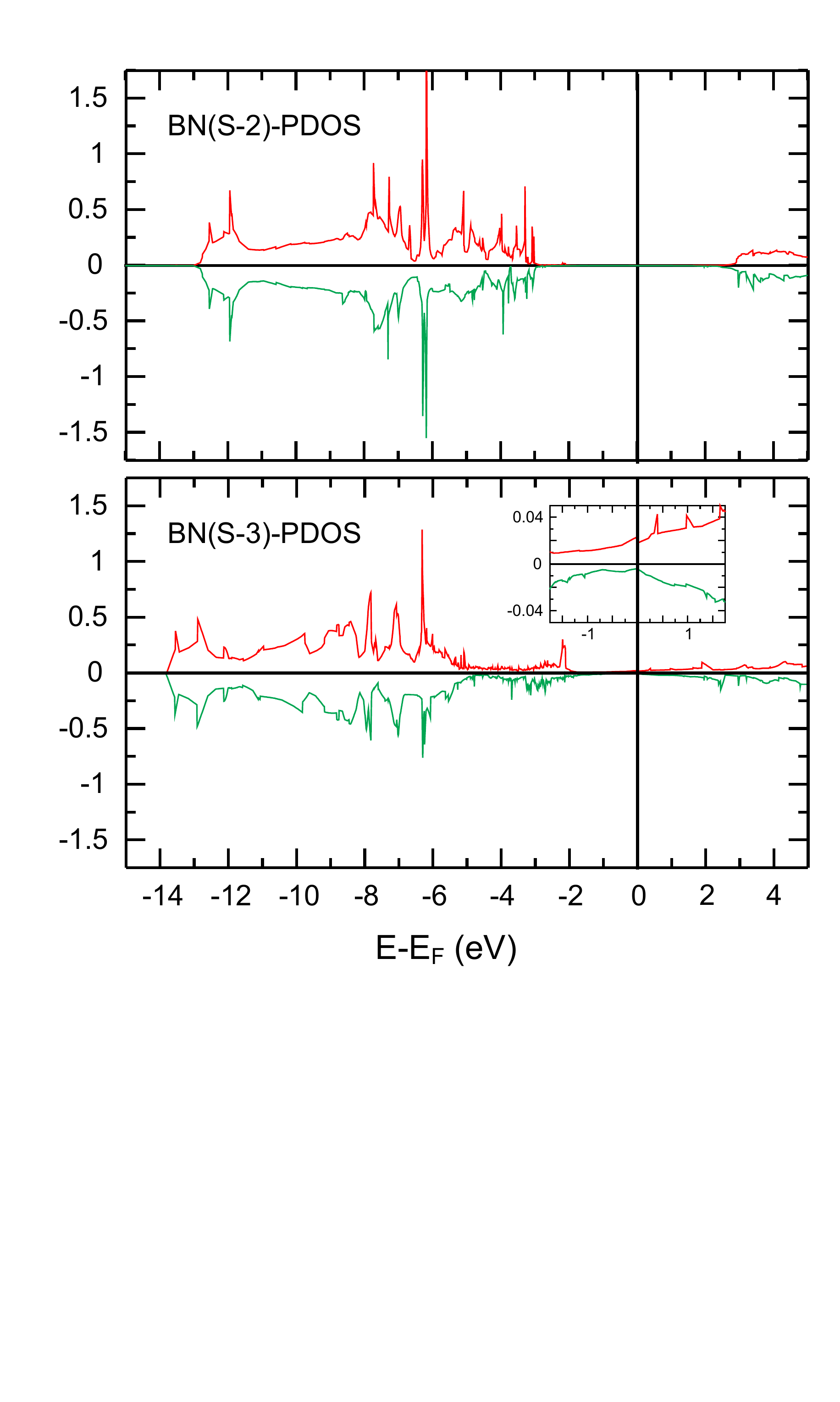}
\end{figure}
\noindent\textbf{Fig.\,S7.} Layer resolved partial density of states (PDOS) for $4$\,ML-h-BN/Ni(111) (spin-up -- top panels; spin-down -- bottom panels).

\clearpage

\begin{figure}
\includegraphics[width=0.48\textwidth]{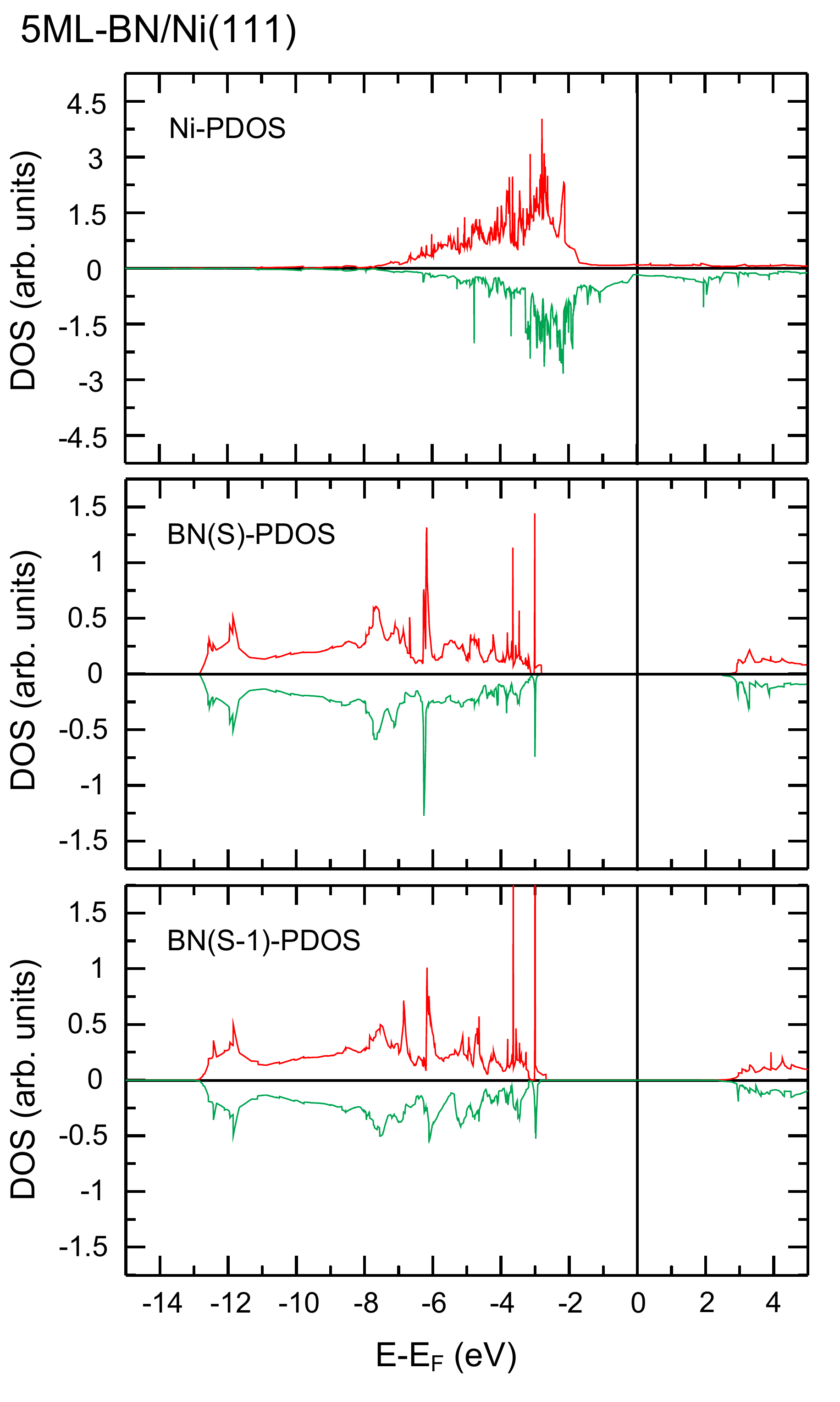}
\hspace{-0.75cm}\includegraphics[width=0.48\textwidth]{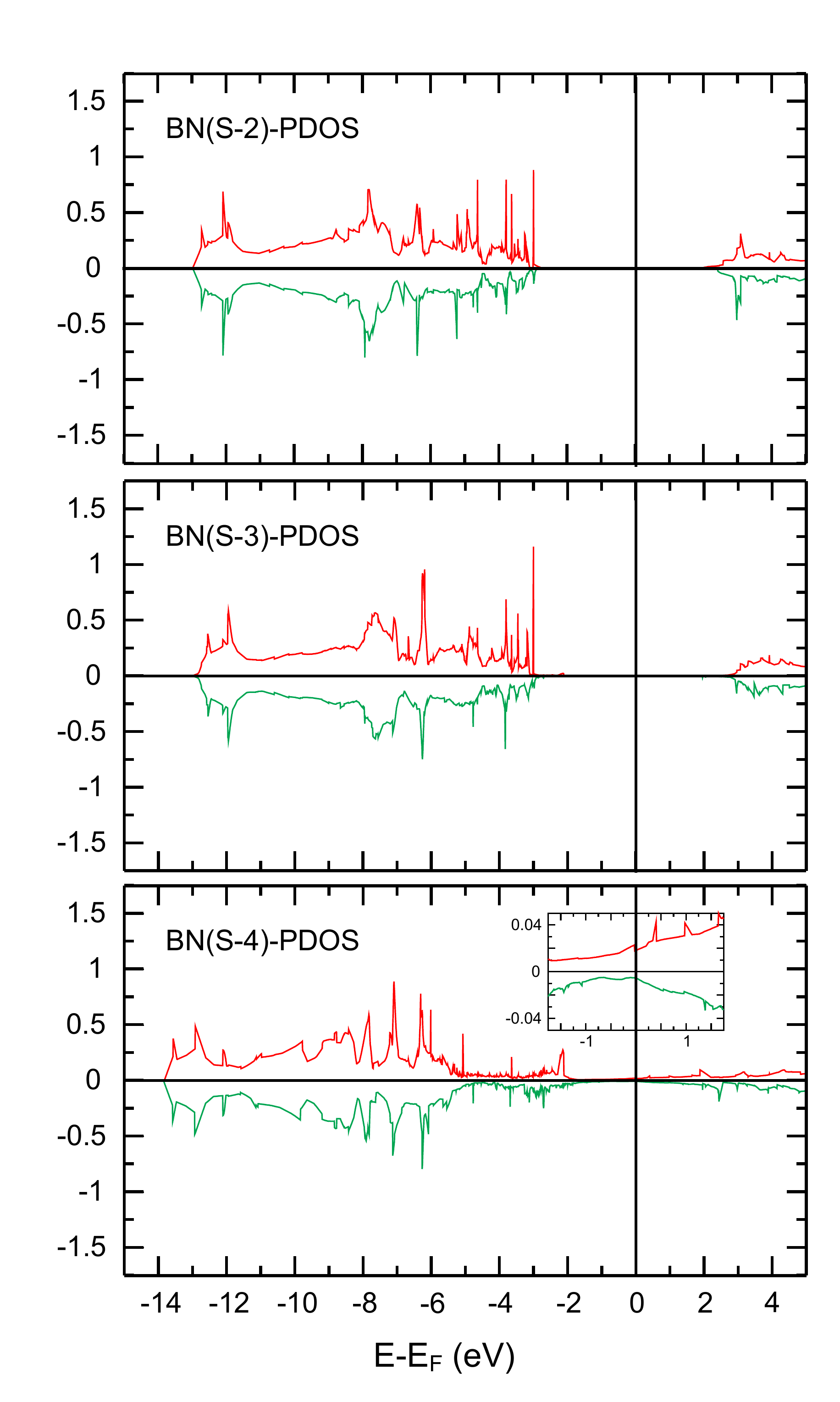}
\end{figure}
\noindent\textbf{Fig.\,S8.} Layer resolved partial density of states (PDOS) for $5$\,ML-h-BN/Ni(111) (spin-up -- top panels; spin-down -- bottom panels).

\end{document}